\documentclass[%
 reprint,
 amsmath,amssymb,
 aps,
prb,
]{revtex4-1}
\usepackage{geometry}                		% See geometry.pdf to learn the layout options. There are lots.
\usepackage{graphicx}				% Use pdf, png, jpg, or eps§ with pdflatex; use eps in DVI mode
\usepackage{amssymb}
\usepackage{amsmath}
\usepackage{romannum}
\usepackage[caption=false]{subfig}
\usepackage{dcolumn}
\usepackage{bm}
\begin{document}
\pagenumbering{arabic}
\setcounter{page}{1}
\preprint{APS/123-QED}

\title{On-site-interaction-induced "Peak-dip-hump" Characteristics with the Mean Field Theory + Perturbation Approach}

\author{Xing Yang}
\affiliation{%
School of Materials Science and Engineering, Guilin University of Electronic Technology, Guangxi Province, China, 541004
\\
 Physics Department, University of Notre Dame, Notre Dame, Indiana, USA, 45665
}%

\date{\today}	

\begin{abstract}

For decades, the difficulty of tackling a strong coupling model with a perturbative approach remained regardless of numerous inquiries. In the current work, a typical mean field theory procedure transforms a strong coupling Hamiltonian into a weak one, which can be solved within a perturbative approach. The Hamiltonian of 2-dimensional free electron gas with Hubbard on-site interaction is calculated as a test of the new procedure. Consequently, the spectral characteristics of the cuprates “peak-dip-hump” are reproduced. The result is compared with the past theoretical investigations, e.g., quantum Monte Carlo method, density matrix renormalization group, tensor network methods, \textit{etc}., which agree well with the spectral curve.

\begin{description}
\item[PACS numbers]
73.23.Hk, 73.40.-c, 73.63.-b, 74.25.-q, 74.78.-w

\end{description}
\end{abstract}

\pacs{73.23.Hk, 73.40.-c, 73.63.-b, 74.25.-q, 74.78.-w}% PACS, the Physics and Astronomy
                             % Classification Scheme.
%\keywords{Suggested keywords}%Use showkeys class option if keyword
                              %display desired
\maketitle
\makeatletter
\newcommand*{\rom}[1]{\expandafter\@slowromancap\romannumeral #1@}
\makeatother

\section{Introduction}
The microscopic mechanism of high-temperature superconductivity has been a heated debate since its discovery in the 1970s. The Hubbard electron-electron on-site interaction is expected to be the critical ingredient for developing high-temperature superconducting theories\cite{Anderson_2002}. The Hubbard model can be solved in a weak coupling limit by the perturbative approach\cite{ZLATIC1992593}\cite{Yoda:1999vx}. While in the intermediate-strong coupling regime, finding the problem’s solution is formidable. Yet some meaningful results by the quantum Monte Carlo method (QMC method)\cite{Bulut:1992wy}\cite{Bulut:1994uu}\cite{Grober:2000wk}\cite{Preuss:1995ti}\cite{Assaad:1996te}, density matrix renormalization group (DMRG)\cite{Yang:2016wl}, tensor network methods\cite{Damme:2021uz}, \textit{etc}., still reach consensus on the spectral curve. These methods are based on different assumptions, which increase the credibility of the theories\cite{Qin_2022}. But there are still some disagreements compared with the experiments. The “peak-dip-hump” in the spectral curve is one of them. 

In the 90s, the perturbative approach was extended to the intermediate-strong coupling regime of the Hubbard model by estimating as many diagrammatic terms as possible. Despite the efforts invested in the issues, the results show uncertainty in explaining the experiments. Subsequently, the QMC method was applied to solve the Hubbard model at half-filling, and the sign problem was absent\cite{Vitali:2016uu}. Recently, the newly invented DMRG approach can obtain the same result by the QMC method but with a higher resolution\cite{Yang:2016wl}. In contrast, the tensor network method efficiently simulates the complex quantum vectors in the Hilbert space\cite{Bohrdt:2020ut}. Noticeably, the dynamic mean-field theory (DMFT) also provides reliable results by mapping the Hubbard model to the impurity model\cite{Lin:2010to}. However, most current practices are suffered from small lattice size, sign problems, or the inability to describe long-range correlations.

By contrast, the variations of the doping and temperature in cuprates exhibit rich physics in the phase diagrams and microscopic mechanisms. For example, electron doping can drive the symmetry of the superconducting order parameter from d-wave to s-wave\cite{Biswas:2002tp}. Recent experiments show the coexistence of the s-wave and d-wave superconductivity in different directions of the bulk cuprates\cite{Zhu:2021uv}. Moreover, the novel “peak-dip-hump” characteristic in Angle-Resolved Photoemission spectroscopy (ARPES) was discovered in YBCO\cite{Chen_2022}\cite{shen1998}\cite{Abanov:1999tj}\cite{Renner:1995tu}, Kagome superconductors\cite{Lou:2022uz}\cite{Nie_2022}, \textit{etc}.

One of the purposes of the paper is to establish an alternative approach to find the solution to the two-dimensional electron system model with the Hubbard-U term. The mean-field theory (MFT) holds when the value of expectations is larger than the fluctuations in the weak coupling limit and the intermediate-strong coupling limit. The philosophy of our procedure is to transform the strong coupling model into a weak coupling model with the mean-field theory and then apply the perturbative approach to the weak coupling model. The first step is to manually select the proper terms with large expectation values and small fluctuations. Secondly, replace the two operators in the two-body strong-coupling interaction term with thermal expectation and neglect its fluctuations. Consequently, the strong coupling model is transformed into a weak coupling one through the operations, which is solvable with the standard perturbative approach. Therefore, this procedure is called the MFT+perturbation approach.

Then we calculate the 2D electron gas with Hubbard on-site interaction by the MFT+perturbation approach. The calculated spectral curve agrees qualitatively with the previous calculation methods. In addition, the new approach can capture the “peak-dip-hump” characteristic in cuprates, which is missing in the previous research. The calculated decay rate of the quasiparticles also qualitatively coincides with Norman’s theory\cite{Norman:1998ub}, although the position of the second singularity has minor distinctions. 

In our recent works\cite{yang2019}\cite{https://doi.org/10.48550/arxiv.2003.07479}, the MFT+Perturbation approach is applied to the 2D electron system at strontium titanate and lanthanum aluminate interface (STO/LAO). The results are consistent with the scanning tunneling spectroscopy, single-electron transistor, and Josephson junction experiments. The STO/LAO interface also shows s-wave unconventional superconductivity. In the paper, we adopted the same Fermi energy, effective mass, \textit{etc}., as the physical quantities of the STO/LAO interface. The system studied in the paper is 2D electron gas with Hubbard on-site interaction, while at the STO/LAO interface, we employed the boson-fermion model. The difference is that the former is gapless at the Fermi level, forbidding the number parity effects, which were first discovered in the single-electron transistor made of aluminum\cite{Randeria:1989aa,Geerligs:1990aa,Tuominen:1992aa,Grabert1992,Lafarge:1993ab,Lafarge:1993aa,Janko:1994aa,PhysRevLett.77.3189}. The number parity effects are not shown in the cuprates in the subsequent experiments. Suppose the electron system in cuprate can be described by the model adopted in the current work. In that case, our calculation supports the observations that the number parity effects are implausible to happen in cuprate.

The paper has four sections. Section I introduces the model of the 2D electron gas with Hubbard on-site interactions. The MFT+Perturbation approach is presented in Section II. The quasiparticle decay rate and density of state are calculated, and the relations with previous works are shown in Section III. We sum up the paper and draw our conclusion in Section IV.

\section{2D electron gas with Hubbard on-site interaction}
The kinetic term of the 2D electron gas is assumed to show a quadratic dispersion. Therefore the Hamiltonian of the 2D electron gas is

\begin{equation}
\hat{T}=\sum_{\mathbf{k},\sigma}{e_\mathbf{k} \hat{c}^\dagger_{\mathbf{k},\sigma} \hat{c}_{\mathbf{k},\sigma}}
\end{equation}

where $e_\mathbf{k}=\frac{\hbar^2 k^2}{2 m^*}-\mu$. $\mu$ is the chemical potential. $\hbar$ is the reduced Planck constant. $m^*$ is the effective mass of the electrons. $ \hat{c}^\dagger_{i,\sigma} (\hat{c}_{i,\sigma})$ is the creation (annihilation) operator of the electrons with momentum $\mathbf{k}$ and spin $\sigma$. The electron gas has a Hubbard on-site interaction

\begin{equation}
\hat{V}=\sum_{i}{U \hat{n}_{i \downarrow} \hat{n}_{i \uparrow}}
\end{equation}

where $U$ is the strength of the on-site interaction and $\hat{n}_{i \sigma}= \hat{c}^\dagger_{i\sigma} \hat{c}_{i\sigma}$. $\hat{c}^\dagger_{\mathbf{k},\sigma} (\hat{c}_{\mathbf{k},\sigma})$ is the creation (annihilation) operator of the electrons with spin $\sigma$ at the lattice site $i$. The interaction in momentum space is

\begin{equation}
\hat{V}=\sum_{i}{\frac{U}{N} \hat{c}^\dagger_{\mathbf{k}_0+\mathbf{q}/2,\uparrow} \hat{c}^\dagger_{-\mathbf{k}_0+\mathbf{q}/2,\downarrow} \hat{c}_{-\mathbf{k}_1+\mathbf{q}/2,\downarrow} \hat{c}_{\mathbf{k}_1+\mathbf{q}/2,\uparrow}}
\end{equation}

where $N$ is the number of the lattice. The total Hamiltonian is

\begin{equation}
\hat{H}=\hat{T}+\hat{V}.
\end{equation}

The Hamiltonian $\hat{H}$ is the same as the BCS Hamiltonian when the momentum $\mathbf{q} = 0$. Hence, the Hamiltonian is the extended BCS Hamiltonian. The extra momentum $\mathbf{q}$ enables the two scattered electrons to have a finite momentum. The terms with $\mathbf{q} \neq 0$ are neglected in the random phase approximation (RPA). In our approach, these terms will produce the decay rate of the quasiparticles and affect the density of states. In this sense, our approach is beyond the RPA.

\section{MFT+Perturbation approach}
The terms with $\mathbf{q} \neq 0$ are viewed as a perturbation to the BCS Hamiltonian. The first step of our approach is to linearize the BCS Hamiltonian with MFT. The transformed kinetic Hamiltonian is

\begin{equation}
\hat{H}_K=\sum_\mathbf{k}
\begin{pmatrix}
\hat{\gamma}_{\mathbf{k}0}^\dagger &\hat{\gamma}_{\mathbf{k}1}^\dagger
\end{pmatrix}
\begin{pmatrix}
\sqrt{e_{\mathbf{k}}^2+\Delta^2} & 0\\
0 &-\sqrt{e_{\mathbf{k}}^2+\Delta^2}
\end{pmatrix}
\begin{pmatrix}
\hat{\gamma}_{\mathbf{k}0} \\
\hat{\gamma}_{\mathbf{k}1}
\end{pmatrix}
\end{equation}

where $\hat{\gamma}_{\mathbf{k}0},\hat{\gamma}_{\mathbf{k}1},\hat{\gamma}_{\mathbf{k}0}^\dagger, \hat{\gamma}_{\mathbf{k}1}^\dagger$ are connected with $\hat{c}_{\mathbf{k}\uparrow},\hat{c}_{\mathbf{-k}\downarrow},\hat{c}_{\mathbf{k}\uparrow}^\dagger, \hat{c}_{\mathbf{-k}\downarrow}^\dagger$ by the Bogoliubov transformation $\hat{\gamma}_{\mathbf{k}0}=u_{\mathbf{k}}\hat{c}_{\mathbf{k}\uparrow}+v_{\mathbf{k}}\hat{c}_{\mathbf{-k}\downarrow}^\dagger$, $\hat{\gamma}_{\mathbf{k}1}=-v_{\mathbf{k}}\hat{c}_{\mathbf{k}\uparrow}+u_{\mathbf{k}}\hat{c}_{\mathbf{-k}\downarrow}^\dagger$. The energy gap $\Delta = \frac{U}{N}\sum_{\mathbf{k}} \hat{c}_{\mathbf{-k}\downarrow} \hat{c}_{\mathbf{-k}\downarrow}^\dagger$. The perturbative terms with $\mathbf{q}\neq 0$ can be reexpressed by the annihilation (creation) operators of the quasiparticles $\hat{\gamma}_{\mathbf{k}0},\hat{\gamma}_{\mathbf{k}1}, \hat{\gamma}_{\mathbf{k}0}^\dagger, \hat{\gamma}_{\mathbf{k}1}^\dagger$. The results contain $\hat{\gamma}\hat{\gamma}\hat{\gamma}\hat{\gamma}$, $\hat{\gamma}^\dagger \hat{\gamma}\hat{\gamma}\hat{\gamma}$, $\hat{\gamma}^\dagger \hat{\gamma}^\dagger \hat{\gamma}\hat{\gamma}$, $\hat{\gamma}^\dagger \hat{\gamma}^\dagger \hat{\gamma}^\dagger \hat{\gamma}$ and $\hat{\gamma}^\dagger \hat{\gamma}^\dagger \hat{\gamma}^\dagger \hat{\gamma}^\dagger$. For simplicity, it is assumed that the amplitude $u_{\mathbf{k}}\gg v_{\mathbf{k}}$ above Fermi level. The Hubbard term will have the form $\frac{U}{N} u u u u \hat{\gamma}^\dagger \hat{\gamma}^\dagger \hat{\gamma}\hat{\gamma}$. Notice that the value of $u_{\mathbf{k}}$ is within a small interval $(\frac{1}{2}, 1)$ and varies slowly with the increase of $|\mathbf{k}|$. As a resullt, $u_{\mathbf{k}}$ can be absorbed by the interaction potential $U$, and $\frac{U}{N} u u u u$ becames $U_T$. In physics, it means that the quasiparticles are subjected to the on-site interaction with a smaller corrected interaction potential compared with the electrons. The transformed Hamiltonian is

\begin{equation}
\hat{H}_T=\hat{H}_K+\hat{H}_V
\end{equation}

where

\begin{equation}
\hat{H}_V=U_T \sum_{\mathbf{k}_0, \mathbf{k}_1,\mathbf{q}\neq0} \hat{\gamma}_{\mathbf{k}_0+\frac{\mathbf{q}}{2}0}^\dagger \hat{\gamma}_{-\mathbf{k}_1+\frac{\mathbf{q}}{2}1} \hat{\gamma}_{-\mathbf{k}_0+\frac{\mathbf{q}}{2}1}^\dagger \hat{\gamma}_{\mathbf{k}_1+\frac{\mathbf{q}}{2}0}.
\end{equation}

The larger terms with $\mathbf{q}=0$ have been diagonalized into the kinetic part $\hat{H}_K$. Therefore the interaction part can be viewed as a perturbative term. Within the perturbation approach, the first and second order diagrams are calculated. In contrast, the self-energies of the first order are real and can be absorbed by the chemical potential and effective mass. As a result, this part has been neglected. The values of the second order self-energies are small. Nevertheless, only one of the second order self-energies can produce imaginary terms. As we later proved, this term is critical to obtain the "peak-dip-hump" characteristics. The second order self-energy is

\begin{equation}
\label{eq:Sigma}
\begin{aligned}
\Sigma(\mathbf{k}, i \omega)= -\frac{U_T^2}{\beta^2 \hbar^4}\int \frac{V d \mathbf{k}_1}{(2 \pi)^2}\int \frac{V d \mathbf{k}_3}{(2 \pi)^2}\sum_{\omega_1, \omega_2}\frac{1}{i \omega-(E_{\mathbf{k}_1}-\mu)/\hbar} \\
\times \frac{1}{i \omega-(E_{\mathbf{k}_3}-\mu)/\hbar}\frac{1}{i (\omega_3+\omega_1-\omega)-(E_{\mathbf{k}_2}-\mu)/\hbar}
\end{aligned}
\end{equation}
where $\beta = k_B T$. $\omega (\mathbf{k}), \omega_1 (\mathbf{k}_1), \omega_3 (\mathbf{k}_3)$ are the frequencies (momenta) of fermionic quasiparticles. $E_\mathbf{k}=\sqrt{e_\mathbf{k}^2+\Delta^2}$ and $\mathbf{k}_2=\mathbf{k}_1+\mathbf{k}_3-\mathbf{k}$. The decay rate of quasiparticles can be obtained by calculaing the imaginary part of the self-energy with analytical continuation $\Gamma = \mathrm{Im}\Sigma(\mathbf{k},i \omega \rightarrow \omega +i \eta^+)$. The real part of the self-energy is small and can be neglected. Concequently, the Green's function of the fermionic quasiparticles is

\begin{equation}
G(\mathbf{k}, \omega)=\frac{1}{\omega-E_{\mathbf{k}}+i\Gamma}.
\end{equation}
 
 The perturbation approach can be applied in the case that $<\hat{H}_K>\gg<\hat{H}_V>$. In RPA, the terms with $\mathbf{q}\neq 0$ are neglected, which produces the Van Hove singularity in the density of states. As a supplement to RPA, the extra terms can induce the decay rate of quasiparticles, the broadening of Van Hove singularity, and the "peak-dip-hump" characteristic.

\section{Quasiparticle decay rate and density of states}

\begin{figure}
  \includegraphics[width=\linewidth]{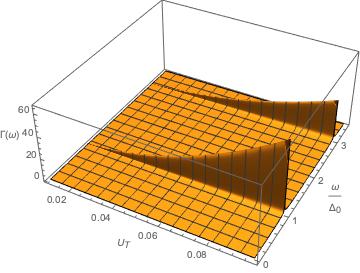}
  \caption{The relation among the decay rate $\Gamma$, frequency of quasiparticles and strength of the interaction potential. The decay rate $\Gamma$ and strength of the interaction potential are in arbitrary units.}
  \label{fig:gamma}
\end{figure}

The decay rate of quasiparticles is calculated from Eq. \ref{eq:Sigma} (see details in Appendix \ref{apd:a} )

\begin{equation}
\Gamma(\omega)=(\frac{V\hbar}{2\pi})^2\frac{\Delta}{\epsilon_F}\frac{U_T^2 m^{*2}}{D_N^2}\frac{N_c^2}{\sqrt{(\omega-\Delta/\hbar)(\omega-3\Delta/\hbar)}}
\end{equation}

where $\epsilon_F$ is the Fermi energy. $D_N$ is the density of states in normal states. $N_c=\int dE \rho(E) \frac{1}{1+\exp(\beta E)}$ and $\rho(E)=\mathrm{Re}\frac{|E|}{\sqrt{E^2-\Delta^2}}$. $N_c$ is the particle number of Bogoliubove quasiparticles. The relation of the decay rate and particle frequency $\omega$ is plotted in Fig \ref{fig:gamma}.

In Fig \ref{fig:gamma}, there are two singularities ($\omega \sim \Delta, \omega \sim 3 \Delta$) of the decay rate $\Gamma$, which induce two zero points of the density of states $D$. The result coincides with norman's theory\cite{Norman:1998ub} based on Eliashberg theory, which also shows two singularities ($\omega \sim \Delta, \omega \sim \Omega_0+ \Delta$) of the decay rate $\Gamma$. $\Omega_0$ is the phonon frequency when the phonon spectral function diverges. In Norman's theory, $\Omega \sim 1.3 \Delta$. The two singularities in our calculations and Norman's theory can produce the "peak-dip-hump" characteristic in cuprates. The calculations of the decay rate also impose restrictions on the temperature regime, at which our calculations are valid. The condition assumed by the calculations is $\epsilon_F\gg\Delta\gg \frac{\hbar^2 |\mathbf{k}|^2}{2m^{*}}-\mu$ and the second inequality demands the energy gap is sufficiently large that the temperature should be lower than a critical temperature $T^*$. Above $T^*$, our calculations will fail. In the two-dimensional electron system (2DES) at the interface of STO/LAO, $T^*\sim 800 \mathrm{mK}$. The Ginzberg criterion indicates the mean field approach will fail when the temperature is above $T^*\sim 800 \mathrm{mK}$ in the 2DES. The demand for the energy gap and the Ginzburg criterion may be different in other systems. However, both conditions can determine the temperature window where the MFT+perturbation approach is valid. Due to the limited number of data sets, the singularities at $U_T\sim 0.02$ are not shown. This may cause an incomplete demonstration of the spectral curve.

The density of states $D$ is calculated by integrating the Green's function over the $k$-space (see details in Appendix \ref{apd:b})

\begin{equation}
\label{eq:dos}
D(\omega)=\int \frac{Vd^2\mathbf{k}}{(2\pi)^2}\frac{1}{\omega-E_\mathbf{k}+i\Gamma}.
\end{equation}

\begin{figure}
  \includegraphics[width=\linewidth]{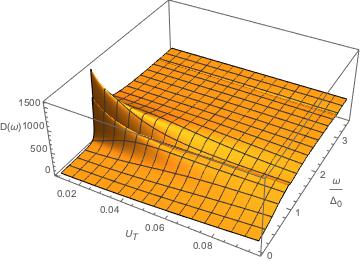}
  \caption{The relation among the density of states, frequency of quasiparticles and strength of the interaction potential. The density of states and strength of the interaction potential are in arbitrary units.}
  \label{fig:dos}
\end{figure}

The result is plotted in Figure \ref{fig:dos}, which shows the "peak-dip-hump" characteristic. The spectral feature is discovered in cuprates, Kagome superconductors\cite{Lou:2022uz}, \textit{etc}. Our result describes an alternative approach to explain the "peak-dip-hump" characteristic in s-wave superconductors with the Hubbard on-site interaction.
The even-odd effect is due to the pairing of electrons in superconductors\cite{Randeria:1989aa,Geerligs:1990aa,Tuominen:1992aa,Grabert1992,Lafarge:1993ab,Lafarge:1993aa,Janko:1994aa,PhysRevLett.77.3189}. The free energy difference of the even and odd states can result in the different lengths of neighboring plateaus in the conductance versus gate voltage curve. In our previous work on the 2DES at the interface of STO and LAO, a finite energy gap is needed for the system to show even-odd effects. The model studied in the present paper is unlikely to give rise to the even-odd effects, which are forbidden by the finite density of states at the Fermi level. The even-odd effects are not found in cuprates experimentally, which are supposed to be described by the presented model. Furthermore, our calculations support the observation that the superconducting materials caused by the on-site interactions between electrons are not expected to show the even-odd effects. Notice that although the superconducting transition temperatures are on the same scale in the two-dimensional electron system at the interface of STO and LAO and the electron system of bulk STO, the superconducting mechanisms are distinct. The former has a preformed-pair state, and the latter has a BCS-like superconducting gap versus the transition temperature curve.

\section{Conclusion}
We solve the two-dimensional electron gas model with the Hubbard on-site interactions employing the MFT+perturbation method. Our approach can capture the "peak-dip-hump" characteristic in cuprates, Kagome superconductors, \textit{etc}. Moreover, it is shown that the electron system with Hubbard-interaction-induced superconductivity is unlikely to show even-odd effects.

The renormalization group approach also adopted the MFT to linearize the Hubbard term, and it focuses more on the evaluation of the superconducting gap and the related phase transition\cite{Gersch_2008}\cite{Gersch_2008,Strack:2008uw,Gersch_2008,Husemann:2009wt,Metzner:2012wd,DUPUIS20211}. Our MFT+perturbation approach mainly investigates the spectral properties of electron systems, which can provide a supplement to the renormalization group approach.

The recent efforts in formulating the non-equilibrium quantum system focus on photon-induced enhancement in superconducting correlations. The exact diagonalization method is one of the main tools, while it is shown that the size effect will incur dramatic changes\cite{Yang_2020}. Our MFT+Perturbation approach may provide an alternative way to solve the time-dependent Hamiltonian since it can give the analytical Green’s function in the strong coupling system.

\begin{acknowledgments}
This work is financially supported by Guangxi Natural Science Foundation under Grant No. AD21220127.
\end{acknowledgments}
\appendix
\section{Appendix A: Calculations of the decay rate}
\label{apd:a}
The calculations of the decay rate $\Gamma$ start from Eq. \ref{eq:Sigma} through obtaining the Matsubara sum of $\omega_1$ and $\omega_3$, which is

\begin{equation}
\Sigma=\int \frac{V^2 d^2\mathbf{k}_1 d^2\mathbf{k}_3}{(2\pi)^4}2\frac{U_T^2}{\hbar^2}\frac{1}{i \omega - (E_{\mathbf{k}_1}-E_{\mathbf{k}_2}+E_{\mathbf{k}_3})/\hbar} \frac{1}{1+\exp{\beta E_{\mathbf{k}_1}}}\frac{1}{1+\exp{\beta E_{\mathbf{k}_3}}}
\end{equation}

The decay rate $\Gamma$ is obtained via analytical continuation. The result is

\begin{equation}
\Gamma=\int \frac{V^2 d^2\mathbf{k}_1 d^2\mathbf{k}_3}{(2\pi)^4}2\frac{U_T^2}{\hbar^2}\delta ( \omega - (E_{\mathbf{k}_1}-E_{\mathbf{k}_2}+E_{\mathbf{k}_3})/\hbar) \frac{1}{1+\exp{\beta E_{\mathbf{k}_1}}}\frac{1}{1+\exp{\beta E_{\mathbf{k}_3}}}.
\end{equation}

With the identity that $\delta(g(x))=\frac{\delta(x-x_0)}{g'(x_0)}$ where $x_0$ is the solution of equation $g(x)=0$, The integrals over the direction angle $\theta_1(\theta_3)$ of the mamentum $\mathbf{k}_1(\mathbf{k}_3)$ are estimated:

\begin{equation}
 \Gamma(|\mathbf{k}|,\omega)=\int \frac{V^2 |\mathbf{k}_1| d|\mathbf{k}_1| |\mathbf{k}_3|d|\mathbf{k}_3|}{(2\pi)^2}\frac{1}{\sqrt{(\omega-\Delta)(\omega-3\Delta)}}\frac{U_T^2}{\hbar^2}\frac{\Delta}{\epsilon_F} \frac{1}{1+\exp{\beta E_{\mathbf{k}_1}}}\frac{1}{1+\exp{\beta E_{\mathbf{k}_3}}}
\end{equation}

where the approximation $|\mathbf{k}_1|, |\mathbf{k}_2|, |\mathbf{k}_3|\sim |\mathbf{k}_F|$ is applied, where $|\mathbf{k}_F|$ is the Fermi wave vector. The integrals over $|\mathbf{k}_1|, |\mathbf{k}_3|$ lead to the final result.

\begin{equation}
\Gamma(\omega)=(\frac{V\hbar}{2\pi})^2\frac{\Delta}{\epsilon_F}\frac{U_T^2 m^{*2}}{D_N^2}\frac{N_c^2}{\sqrt{(\omega-\Delta/\hbar)(\omega-3\Delta/\hbar)}}.
\end{equation}

\section{Appendix B: Calculations of the density of states}
\label{apd:b}
Starting from Eq. \ref{eq:dos}, the residue theorem can be applied to obtain the analytical form of the density of states. For simplicity, the chemical potential is neglected. Two solutions of the momentum $k$ are in the right upper half-plane of the complex plane of $k$, which will be counted. If define $a=\omega^2-\Gamma^2-\Delta^2, b=2 \Gamma \omega$, four different cases are

  \[
                \left\{
                \begin{aligned}
                  a>0, b>0...................\romannum{1}~~\\
                  a>0, b<0...................\romannum{2}~\\
                  a<0, b>0...................\romannum{3}\\
                  a<0, b<0...................\romannum{4}.
                \end{aligned}
              \right.
  \]

For the first case, two solutions of the momentum $k$ in the right upper half-plane are

                \begin{eqnarray}
                   k_1=\sqrt{2m^*}R^{\frac{1}{4}}\exp{\frac{\theta i}{4}}\\
                   k_2=i\sqrt{2m^*}R^{\frac{1}{4}}\exp{\frac{-\theta i}{4}}.
                \end{eqnarray}

where $\theta=\mathrm{Arctan}\frac{b}{a}, R=\sqrt{a^2+b^2}$. For the second case,

                \begin{eqnarray}
                   k_1=\sqrt{2m^*}R^{\frac{1}{4}}\exp{\frac{-\theta i}{4}}\\
                   k_2=i\sqrt{2m^*}R^{\frac{1}{4}}\exp{\frac{\theta i}{4}}.
                \end{eqnarray}

For the third case,

                \begin{eqnarray}
                   k_1=\sqrt{2m^*}R^{\frac{1}{4}}\exp{\frac{(\theta+\pi) i}{4}}\\
                   k_2=i\sqrt{2m^*}R^{\frac{1}{4}}\exp{\frac{(-\theta-\pi) i}{4}}.
                 \end{eqnarray}

For the fourth case,

                \begin{eqnarray}
                   k_1=\sqrt{2m^*}R^{\frac{1}{4}}\exp{\frac{(-\theta+\pi) i}{4}}\\
                   k_2=i\sqrt{2m^*}R^{\frac{1}{4}}\exp{\frac{(\theta-\pi) i}{4}}.
                  \end{eqnarray}

  After applying Jordan's lemma and the residue theorem, we obtain the density of states, for case one,

\begin{equation}
D(\omega)=\frac{m^*V}{2\pi}(\frac{\omega }{\sqrt{R}}cos(\frac{\theta}{2})+\frac{\Gamma}{\sqrt{R}}sin(\frac{\theta}{2})).
\end{equation}

For case two,
\begin{equation}
D(\omega)=-\frac{m^*V}{2\pi}(\frac{\omega }{\sqrt{R}}cos(\frac{\theta}{2})+\frac{\Gamma}{\sqrt{R}}sin(\frac{\theta}{2})).
\end{equation}

For case three,
\begin{equation}
D(\omega)=\frac{m^*V}{2\pi}(\frac{\omega }{\sqrt{R}}cos(\frac{\theta+\pi}{2})+\frac{\Gamma}{\sqrt{R}}sin(\frac{\theta+\pi}{2})).
\end{equation}

For case four,
\begin{equation}
D(\omega)=-\frac{m^*V}{2\pi}(\frac{\omega }{\sqrt{R}}cos(\frac{\theta-\pi}{2})+\frac{\Gamma}{\sqrt{R}}sin(\frac{\theta-\pi}{2})).
\end{equation}
\bibliographystyle{apsrev4-1}
\bibliography{refs.bib}
\end{document}